\documentclass[twocolumn,]{aastex63}

\usepackage[T1]{fontenc}

\DeclareRobustCommand{\VAN}[3]{#2}
\let\VANthebibliography\thebibliography
\def\thebibliography{\DeclareRobustCommand{\VAN}[3]{##3}\VANthebibliography}

\received{December 6, 2020}
\revised{January 21, 2021}
\accepted{January 22, 2021}
\submitjournal{ApJ Letters}

\usepackage{graphicx}	
\usepackage{amsmath}	
\usepackage{amssymb}	
\usepackage{enumitem}

\newcommand{\be}{\begin{equation}}
\newcommand{\ee}{\end{equation}}

\shorttitle{Star Formation Distributions in Fiery Cores}
\shortauthors{M. E. Orr et al.}

\begin{document}
\label{firstpage}

\title{Fiery Cores: Bursty and Smooth Star Formation Distributions across Galaxy Centers in Cosmological Zoom-in Simulations}
\correspondingauthor{Matthew Orr}
\email{matt.orr@rutgers.edu}

\author[0000-0003-1053-3081]{Matthew E. Orr}
\affiliation{TAPIR, Mailcode 350-17, California Institute of Technology, Pasadena, CA 91125, USA}
\affiliation{Department of Physics and Astronomy, Rutgers University, 136 Frelinghuysen Road, Piscataway, NJ 08854, USA}
\affiliation{Center for Computational Astrophysics, Flatiron Institute, 162 Fifth Avenue, New York, NY 10010, USA}

\author[0000-0003-0946-4365]{H Perry Hatchfield}
\affiliation{University of Connecticut, Department of Physics, 196A Auditorium Road, Unit 3046, Storrs, CT 06269 USA}

\author[0000-0002-6073-9320]{Cara Battersby}
\affiliation{University of Connecticut, Department of Physics, 196A Auditorium Road, Unit 3046, Storrs, CT 06269 USA}

\author[0000-0003-4073-3236]{Christopher C. Hayward}
\affiliation{Center for Computational Astrophysics, Flatiron Institute, 162 Fifth Avenue, New York, NY 10010, USA}

\author[0000-0003-3729-1684]{Philip F. Hopkins}
\affiliation{TAPIR, Mailcode 350-17, California Institute of Technology, Pasadena, CA 91125, USA}

\author[0000-0003-0603-8942]{Andrew Wetzel}
\affiliation{Department of Physics \& Astronomy, University of California, Davis, CA 95616, USA}

\author[0000-0003-4826-9079]{Samantha M. Benincasa}
\affiliation{Department of Physics \& Astronomy, University of California, Davis, CA 95616, USA}

\author[0000-0003-3217-5967]{Sarah R. Loebman}
\altaffiliation{Hubble Fellow}
\affiliation{Department of Physics \& Astronomy, University of California, Davis, CA 95616, USA}

\author[0000-0001-6113-6241]{Mattia C. Sormani}
\affiliation{Universit\"{a}t Heidelberg, Zentrum f\"{u}r Astronomie, Institut f\"{u}r Theoretische Astrophysik, Albert-Ueberle-Str. 2, 69120 Heidelberg, Germany}

\author[0000-0002-0560-3172]{Ralf S. Klessen}
\affiliation{Universit\"{a}t Heidelberg, Zentrum f\"{u}r Astronomie, Institut f\"{u}r Theoretische Astrophysik, Albert-Ueberle-Str. 2, 69120 Heidelberg, Germany}
\affiliation{Universit\"{a}t Heidelberg, Interdisziplin\"{a}res Zentrum f\"{u}r Wissenschaftliches Rechnen, Im Neuenheimer Feld 205, 69120 Heidelberg, Germany}

\begin{abstract}
We present an analysis of the $R\lesssim 1.5$~kpc core regions of seven simulated Milky Way mass galaxies, from the FIRE-2 (Feedback in Realistic Environments) cosmological zoom-in simulation suite, for a finely sampled period ($\Delta t = 2.2$ Myr) of 22 Myr at $z \approx 0$, and compare them with star formation rate (SFR) and gas surface density observations of the Milky Way's Central Molecular Zone (CMZ). Despite not being tuned to reproduce the detailed structure of the CMZ, we find that four of these galaxies are consistent with CMZ observations at some point during this 22 Myr period.  The galaxies presented here are not homogeneous in their central structures, roughly dividing into two morphological classes; (a) several of the galaxies have very asymmetric gas and SFR distributions, with intense (compact) starbursts occurring over a period of roughly 10 Myr, and structures on highly eccentric orbits through the CMZ, whereas (b) others have smoother gas and SFR distributions, with only slowly varying SFRs over the period analyzed.  In class (a) centers, the orbital motion of gas and star-forming complexes across small apertures ($R \lesssim 150$pc, analogously $|l|<1^\circ$ in the CMZ observations) contributes as much to tracers of star formation/dense gas 
appearing in those apertures, as the internal evolution of those structures does.  These asymmetric/bursty galactic centers can simultaneously match CMZ gas \emph{and} SFR observations, demonstrating that time-varying star formation can explain the CMZ's low star formation efficiency.

\end{abstract}



\section{Introduction}

Within the context of empirical star formation laws, galaxy centers often exhibit particularly extreme and peculiar properties. From observations on scales averaging over entire galaxies down to those of $\sim$100 pc, the star formation rate (SFR) scales with the surface density of molecular gas as a power law relationship known as the Kennicutt-Schmidt law \citep[KS,][]{Schmidt1959,Kennicutt1998}. Moreover, the presence of high density ($> 10^4$ cm$^{-3}$) gas seems to strongly predict the star formation rate on the scale of individual molecular clouds  \citep{Lada2010,lada_Star_2012a}. Some galaxy centers host nuclear starbursts (e.g.\ NGC253, \citealt{leroy_forming_2018}), whereas the Milky Way's central region, known as the Central Molecular Zone (CMZ) appears to be under-producing stars relative to its dense gas content, with surveys finding a SFR 10-100 times lower than that predicted by contemporary theory (\emph{e.g.}, \citealt{immer_recent_2012, Longmore2013}). This deficiency has motivated many studies of the star forming properties of molecular clouds in this extreme environment (\citealt{rathborne_g0.253+0.016:_2014, Ginsburg2018, walker_Star_2018, barnes_Young_2019,henshaw_Brick_2019}).

While there is resilient evidence for this star formation deficiency in the Milky Way's CMZ \citep{barnes_star_2017}, there are still signs of ongoing and previously more intense past star formation episodes within the past 2-6 Myr (\emph{e.g.}, \citealt{Liermann2012, Lu2013, Clark2018}). It has been suggested that the CMZ may have previously been in a more vigorous state of star formation, perhaps similar to other galaxies with nuclear starbursts \citep{Kruijssen2014a, krumholz_Dynamical_2017, Sormani2020}. Arguments that the CMZ has a low star formation efficiency per dense gas mass often presuppose that the CMZ is in equilibrium, and the time evolution of galactic centers is difficult to study using Milky Way observations alone. 

Although limited to probing scales from 10-100~pc, extragalactic galaxy center studies have measured star formation and gas in MW-mass galaxy centers across a range of conditions \citep[][among others]{Casasola2015, Gallagher2018}.  These observations have highlighted the variety of conditions under which star formation occurs in galaxy centers, further suggesting that large variations ($\sim$dex) in SFRs and gas surface densities naturally arise \citep{Leroy2013} in these extreme environments.


Exploring the nature of star formation in galactic centers requires detailed modeling of star formation and feedback processes \citep[\emph{e.g.},][]{armillotta_life_2019}, as well as a self-consistent picture of gas dynamics in the full context of galactic structure and evolution \citep[\emph{e.g.},][]{tress_Simulations_2020,sormani_Simulations_2020}. Cosmological zoom-in simulations have begun to meet these physics requirements, and now have adequate spatial/mass resolution to follow the multiphase turbulent ISM, and capture the cosmological context of Milky Way-like galaxies \citep[\emph{e.g.},][]{Hopkins2014, Hopkins2018:fire}.  In particular, work within the FIRE collaboration has been able to explore star formation in the context of galactic disks, cloud lifetimes, and SMBH--gas dynamics connections \citep{Angles-Alcazar2017, Orr2018, Orr2020, Benincasa2020, Gurvich2020}.  Recent work by \citet{Sanderson2020} has compared several galaxies in the FIRE-2 suite in detail to properties of the MW.

In this letter, we compare the centers of seven FIRE-2 galaxies \citep{Wetzel2016, Hopkins2018:fire}, all approximately Milky Way-mass spirals, with Milky Way CMZ and extragalactic observations.  These simulations have $z=0$ SFRs of $3 - 10$ M$_\odot$/yr, which is more typical of $L^\star$ galaxies compared the MW (our Galaxy appears to be an outlier to lower SFR, \citealt{Longmore2013}).  Specifically, we map the centers of the simulated galaxies at high spatial resolution to understand the effects of dynamical evolution and feedback over a short ($\sim 22$ Myr) timescale on proxies for SFR and gas surface density tracers, and subsequent interpretations of star formation activity in their central regions.

\begin{figure*}
	\centering
	\includegraphics[width=0.97\textwidth]{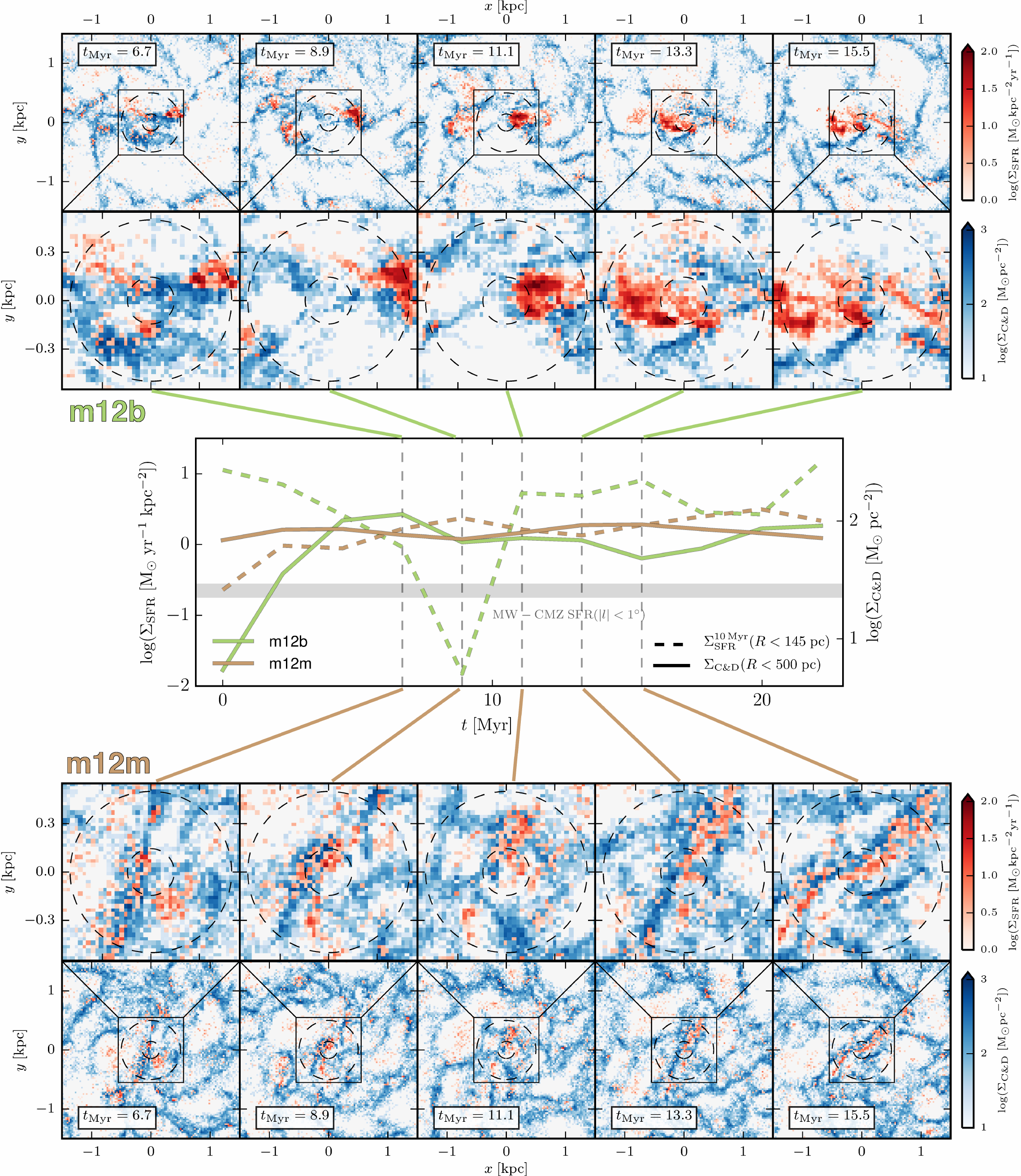}
	\caption{Face-on central regions of two FIRE-2 spiral galaxies at five snapshots in time (advancing right, $\Delta t \approx 2.2$ Myr): \textbf{m12b} (top subfigure) and \textbf{m12m} (bottom subfigure), cold and dense gas (C\&D, $T < 500$~K and $n_{\rm H} > 1$ cm$^{-3}$) in blues with $10$ Myr-averaged SFR (reds) overlaid, with 25 pc pixel sizes. Outermost rows show $3$ kpc regions, with inner rows showing $1.1$ kpc zoom-ins.  Zoom-in panels show two apertures, with $R= 145, 500$ pc (inner, outer dashed lines respectively).  Middle panel shows the time evolution of SFRs within $R <145$ pc (dashed lines) and gas surface densities within $R <500$ pc (solid lines). MW CMZ SFR estimate with uncertainty ($|l|<1^\circ$ from \citealt{Longmore2013}) plotted as horizontal grey-shaded band.  Despite having similar gas surface densities on $R < 500$ pc scales (modulo \textbf{m12b} lacking C\&D gas within $\sim$300 pc for the first $\sim$5 Myr), the two galaxies have markedly different star formation in their centers, with \textbf{m12b} (green lines) having bursty, intense star formation as opposed to the smoother cirrus of star formation seen in \textbf{m12m} (brown lines).  The large SFR variation in \textbf{m12b} essentially corresponds to the evolution and physical motion of a single massive star-forming region.}
	\label{fig:twogals}
\end{figure*}

\section{Methods}

We analyze the central regions the seven Milky Way/Andromeda-mass spiral galaxies from the `standard physics' Latte suite of FIRE-2 simulations introduced in \citet{Wetzel2016} and \citet{Hopkins2018:fire}. The spatially resolved properties of the gas surface densities, velocity dispersions, and SFRs across the disks of these galaxies have been studied in detail in \citet{Orr2020}. This work makes use of 11 snapshots finely spaced in time ($\Delta t \approx 2.2$ Myr) at $z \approx 0$ for each of the simulations.  A brief summary of the $z \approx 0$ global properties of the galaxy simulations are included in Table~1 of \citet{Orr2020}.  

The simulations analyzed here all have minimum baryonic particle masses of $m_{b,min} = 7100$ M$_\odot$, minimum adaptive force softening lengths $<$1~pc, and a 10 K gas temperature floor.  With adaptive softening lengths, we note that the median softening length within the disk in the runs at $z=0$ is $h\sim 20-40$ pc (at a $n\sim 1$ cm$^{-3}$), with the dense turbulent disk structures having necessarily shorter softening lengths. The aperture sizes considered in this work are $145-500$ pc\footnote{For comparison with CMZ observations, a physical radial extent of $R\lesssim 145$~pc corresponds to $|l|\lesssim 1^\circ$ in Galactic longitude, assuming a distance of $d\approx 8.2$~kpc.}, and so are well above the minimum resolvable scales in the simulations.

Importantly, for discussion here: star formation in the simulations occurs on a free-fall time in gas which is dense ($n >10^3$ cm$^{-3}$), molecular (per the \citealt{Krumholz2011} prescription), self-gravitating (viral parameter $\alpha_{\rm vir} < 1$) and Jeans-unstable below the resolution scale. Once these requirements are met, the SFR at the particle scale is assumed to be: $\dot\rho_\star = \rho_{\rm H_2}/t_{\rm ff}$ (\emph{i.e.}, 100\% efficiency per free-fall time). Star particles are treated as single stellar populations, with known age, metallicity, and mass.   Feedback from supernovae, stellar mass loss (OB/AGB-star winds), photoionization and photoelectric heating, and radiation pressure are explicitly modeled.  These simulations do not include any supermassive black holes (SMBHs), and accordingly do not have any feedback associated with BH accretion, nor do they include cosmic rays or other MHD physics. Detailed descriptions of these physics and their implementation can be found in \citet{Hopkins2018:fire}.

We produce mock observational maps from the snapshots using the same methods as \citet{Orr2018} and \citet{Orr2020}: we project the galaxies face-on according to their stellar angular momentum vector (including star particles out to 20~kpc) and bin star particles and gas elements into square pixels with side-lengths (\emph{i.e.}, ``pixel sizes'') 25~pc. The maps analyzed here are 3~kpc on a side, and integrate gas and stars within $\pm 1.5$ kpc of the galactic mid-plane. 

We generate a proxy for observational measures of recent SFRs by calculating the 10 Myr-averaged SFR in the pixels.  We do this by summing the mass of star particles with ages less than 10 Myr, and correcting for mass loss from stellar winds and evolutionary effects using predictions from {\scriptsize STARBURST99} \citep{Leitherer1999}.  This time interval was chosen for its approximate correspondence with the timescales traced by recombination lines like H$\alpha$ \citep{Kennicutt2012}\footnote{A direct comparison to observations, by post-processing the snapshots to explicitly model H$\alpha$, would make for a more accurate modeling, but accounting for, \emph{e.g.}, dust and other complexities, is beyond this letter's scope, where we wish to focus on the ``true'' SFRs and their spatial distributions. Work by \citet{Velazquez2020} has suggested that shorter ($\sim$5 Myr) timescales more accurately trace H$\alpha$ emission.  We use a slightly longer averaging window such that the simulations well-resolve SFR estimates of the CMZ over these timescales within $R<145$~pc (where $\sim 10$ young star particles corresponds to a measured SFR of $\sim$0.1 M$_\odot$ yr$^{-1}$ kpc$^{-2}$), and are more conservative in our sensitivity to SFR variability.}.  To compare with gas observations, we calculate column densities for the ``cold and dense'' gas ($\Sigma_{\rm C\&D}$ throughout) with $T < 500$~K and $n_{\rm H} > 1$ cm$^{-3}$.  This gas reservoir taken as a proxy for the cold molecular gas in the simulations following the methodology of \citet{Orr2020}, and ought roughly to correspond with gas traced by cold dust or CO observations in the CMZ.


\section{Results}
\begin{figure*}
	\centering
	\includegraphics[width=0.84\textwidth]{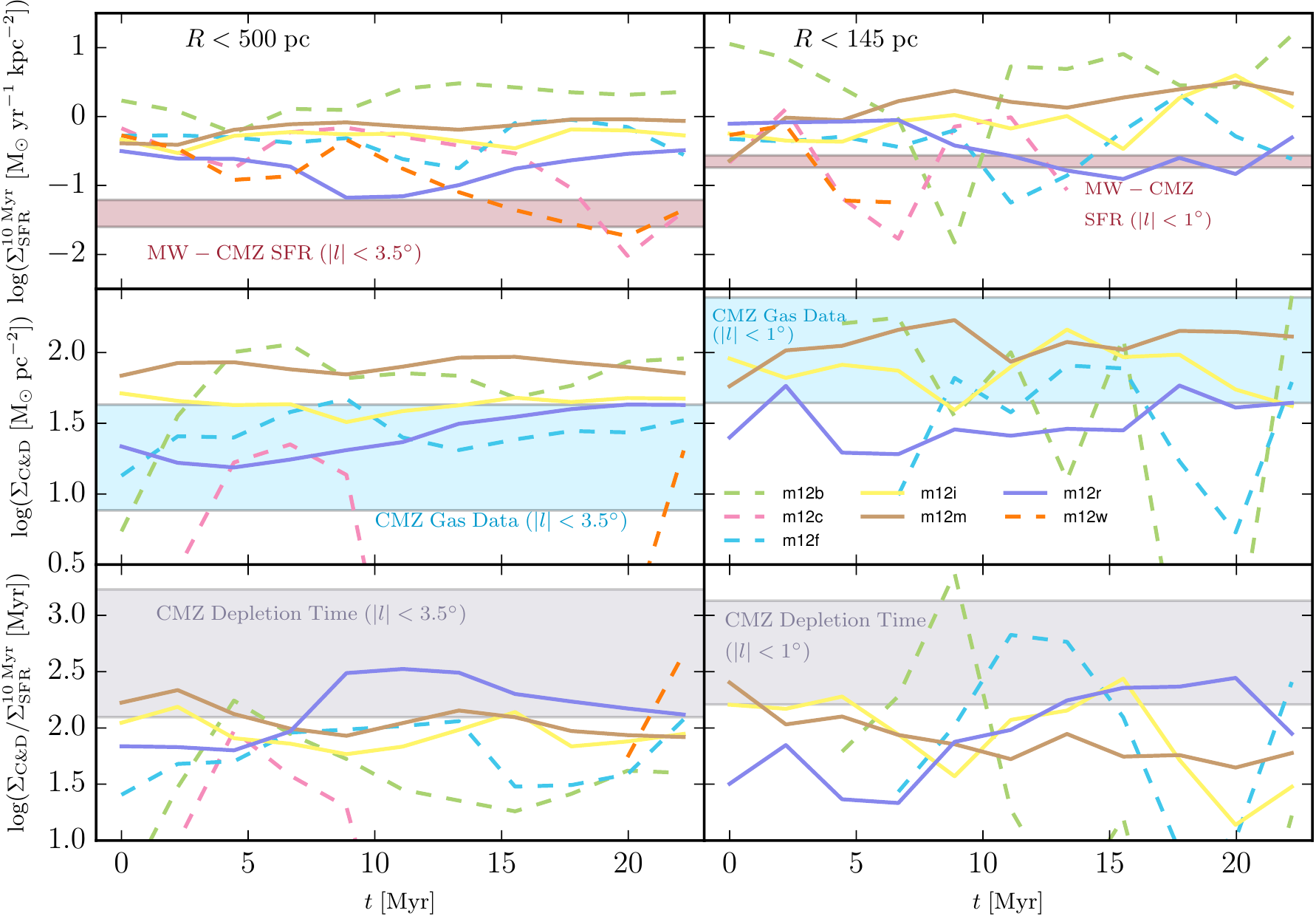}
	\caption{SFRs and cold \& dense (C\&D) gas surface densities in central regions of seven FIRE-2 spiral galaxies (colored lines; ``class (a)/(b)'' plotted with dashed/solid lines, respectively), for $R < 500$ (left column) and $R < 145$ pc (right column) apertures, as a function of time near $z \approx 0$ ($\Delta t \approx 2.2$ Myr, rightmost edge being $z=0$).  Shaded bands indicate SFR and gas surface density observations, with uncertainty, of the CMZ from \citet{Longmore2013} and \citet{Mills2017}, respectively. Depletion times ($\Sigma_{\rm C\&D}/\Sigma_{\rm SFR}^{\rm 10 \; Myr}$) are also presented, in the same style; these CMZ depletion times are produced by combining \citet{Longmore2013} and \citet{Mills2017} data.  SFRs evolve more smoothly in all galaxies in larger apertures ($R < 500$ pc), and the variance in SFRs or gas surface density increases with smaller apertures.  However, in the simulations, two central molecular zone classes appear to exist on $R<145$~pc scales: galaxies like \textbf{m12b} and \textbf{m12c} with very asymmetric gas distributions and dramatic starbursts on $\sim 10$ Myr timescales, ``class (a)''; and galaxies like \textbf{m12i} and \textbf{m12m} typifying smoother (though still with non-trivial fluctuations) SFR and gas distributions in their centers, ``class (b)'' (see, \textbf{m12b} and \textbf{m12m} in Fig.~\ref{fig:twogals} as examples of classes (a) and (b), respectively).  Despite temporal and spatial variance, many of the FIRE galaxies are consistent with MW CMZ observations \emph{at some point} in this time-window.}
	\label{fig:GASnSFRpanel}
\end{figure*}

In this sample of only seven FIRE-2 Milky Way analogues, there is a surprising variety of conditions in their centers. As an example of properties seen in the galaxy centers, Figure~\ref{fig:twogals} shows 8.8 Myr of two of the FIRE-2 galaxies (\textbf{m12b} and \textbf{m12m}), and how their SFRs and gas surface densities evolve within their central $\sim 500$~pc (zoomed insets).  Despite having similar masses of cold gas in their galactic centers, the two simulations have morphologically distinct central regions, in terms of their cold gas distributions and star-forming complexes.  Of the two galaxies in Figure~\ref{fig:twogals}, only \textbf{m12b} is able to match observational estimates using YSO counts and HII regions of the CMZ SFR within $|l| <1^\circ$ \citep{Longmore2013}, specifically this is seen to occur during an inter-starburst period (at $t=8.9$ Myr, center-left column).  At least four of the FIRE galaxies (\textbf{m12b}, \textbf{m12f}, \textbf{m12m}, \& \textbf{m12r}) match CMZ SFR \emph{and} gas surface density properties concurrently \emph{at some time} in this $22$ Myr period.  

\subsection{Two Morphological Classes of FIRE-CMZs}\label{sec:morph}
The FIRE galaxies presented here cover a range of morphologies in their centers because these simulations were \emph{not designed to match the detailed structural properties of our Galactic center}. Within the sample of seven galaxies, we see two distinct classes of central morphology in their \emph{fiery cores} (gas and star formation distributions within $R \approx 1.5$ kpc):  

\begin{enumerate}[label=(\alph*)]
    \item ``Asymmetric/Bursty'' (\textbf{m12b}, \textbf{m12c}, \textbf{m12f}, \& \textbf{m12w}): Large, asymmetric gas clouds and star-forming complexes are seen. Star formation is concentrated in intense starbursts whose feedback dramatically shapes the local gas environment (see \textbf{m12b}, upper subfigure of Fig.~\ref{fig:twogals}). Two simulations falling in this category (\textbf{m12b} \& \textbf{m12f}) simultaneously match the MW CMZ gas \emph{and} SFR measurements. The two others (\textbf{m12c} \& \textbf{m12w}) do not simultaneously have SFR \emph{and} dense gas tracers within the central 145~pc \emph{at any point in this time window}. 
    
    \item ``Smooth'' (\textbf{m12i}, \textbf{m12m}, \& \textbf{m12r}): Gas and star formation is smoothly distributed within the galactic centers, with clear feeding of gas into center, and a cirrus of star formation (see \textbf{m12m}, lower subfigure of Fig.~\ref{fig:twogals}). Feedback events do not dramatically alter the local gas environment, as the feedback is relatively dispersed across their centers. 
\end{enumerate}

Interestingly, none of the galaxies here exhibit the ring structures, presumed to be long-lived, seen by studies of the central regions of other spiral galaxies and the MW CMZ \citep{Kormendy2004, Molinari2011}. 
We note the lack of strong bars in the centers of \emph{any} of these FIRE galaxies \emph{at this time} (\textbf{m12m} did develop a strong bar around $z\approx 0.2$, but it does not survive to $z=0$; \citealt{Debattista2019})
; without the presence of bars in these simulations at $z=0$, we cannot speak to the dynamical importance of bars in producing central galactic environments similar to the MW CMZ. Other work has highlighted the potential impacts of bars in funneling gas core-ward and forming rings \citep{Sormani2015, Sormani2018}.
However, bar-induced effects would likely push these simulated galactic centers towards more asymmetric states, supporting the idea that bursty, rather than steady-state, star formation is necessary to explain MW CMZ observations.  Specifically, \citet{Sormani2018} showed that the gas flow in a barred potential naturally becomes turbulent and asymmetric, even in the absence of any type of stellar feedback.  And so, we leave it to future work to investigate the gas flows driven in FIRE galaxies by the bars that form at higher redshift.

The structures in the centers of the FIRE galaxies appear to be fairly transient in nature, existing for $\lesssim 10$ Myr (similar to the GMC lifetimes seen in these simulations more broadly by \citealt{Benincasa2020}).  We should be clear: spiral structures do exist core-ward in these simulations (see the clear presence of spiral arms in both \textbf{m12b} and \textbf{m12m} in Fig.~\ref{fig:twogals}).  In the case of the class (a) morphologies, the central ($R< 300$ pc) gas structures are on visibly eccentric orbits through their CMZs (similar to MW CMZ orbital modeling by \citealt{Kruijssen2015}), and are disrupted 
by intense feedback following starbursts.  Previous work with FIRE by \citet{Torrey2017} showed that star formation--feedback instabilities in galactic centers can arise when the local dynamical time becomes shorter than the feedback timescale.   The class (b) morphologies have smoother gas distributions in their centers, and with the gentler, more dispersed feedback from diffuse star formation, are not as strongly disrupted.  Owing perhaps to their smoother gas distributions, less of the gas is on very eccentric orbits ($\sim5-10$\% of gas having $v_{\rm in-plane} > \sqrt{2}v_c$, vs. up to $\sim 20-40$\% in the class (a) centers) and, to an extent, the structures are clearer continuations of spiral arms down to their centers.  The difference between the two classes may, in the case of these FIRE galaxies, arise from more or less violent recent merger histories/interactions with (smaller) galaxies, with the class (b) galaxies having more quiescent recent histories. To wit, as shown in \citet{Garrison-Kimmel2018}, \textbf{m12i} and \textbf{m12m} have not experienced any notable head-on major mergers. 


One caveat to the discussion regarding these morphologies is the lack of SMBHs in these simulations.  Work by \citet{Angles-Alcazar2017} has investigated the influence of SMBHs on their immediate environments, and their ability to disrupt gas structures while they are actively accreting, may disallow the smooth central gas distributions within $\sim$100 pc in class (b).  And so, class (a) galactic centers may be the more realistic central galactic environments. 

\subsection{Matching FIRE-CMZs with the Milky Way CMZ and External Galaxies}

\begin{figure}
	\centering
	\includegraphics[width=0.47\textwidth]{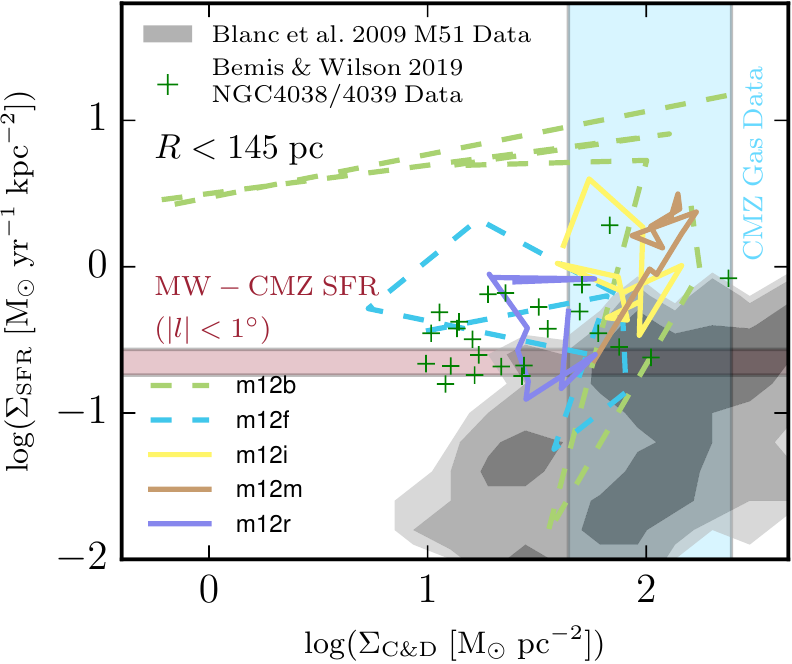}
	\caption{KS relation in central regions of the five FIRE-2 spiral galaxies (colored lines: ``asymmetric'' centers plotted with dashed lines, ``smooth'' centers with solid lines) that \emph{simultaneously} have SFR and dense gas tracers within the $R< 145$ pc aperture, as a function of time near $z \approx 0$ ($\Delta t \approx 2.2$ Myr). CMZ SFR and gas estimates, with uncertainty, (\citealt{Longmore2013} and \citealt{Mills2017}) plotted as horizontal and vertical shaded bands, respectively, and spatially resolved KS observations ($\sim 170$ pc \& $\sim 675$, respectively) of M51 \citep[][their $X_{\rm CO}$ adjusted to be consistent with MW value]{Blanc2009} and the Antennae Galaxies \citep[NGC 4038/9;][]{Bemis2019} plotted in greyscale contours and with green `+'s, respectively. 
	The central regions of some galaxies remain fairly stable in KS-space over $22$ Myr (\emph{e.g.}, \textbf{m12m}), whereas others (\emph{e.g.}, \textbf{m12b}) vary by upwards of a dex in both SFR and $\Sigma_{\rm C\&D}$. Four FIRE galaxies (\textbf{m12b}, \textbf{m12f}, \textbf{m12m}, \& \textbf{m12r}) overlap with the CMZ SFR estimate at various times.} 
	\label{fig:KS_1deg}
\end{figure}

Fig.~\ref{fig:GASnSFRpanel} shows the evolution over 22 Myr of the SFR and cold \& dense (C\&D) gas surface densities, and derived depletion times, within $R = 500$ and $145$ pc apertures (corresponding roughly to CMZ observations within $|l| \lesssim 3.5^\circ$ and $1^\circ$, with SFRs taken from \citealt{Longmore2013} and gas from \citealt{Mills2017}) for the FIRE galaxies.  Averaging over larger scales reduces the degree of scatter seen in SFR and gas surface density for each galaxy.  However, as discussed in \S\ref{sec:morph}, the time evolution of $\Sigma_{\rm SFR}$ and $\Sigma_{\rm C\&D}$ alone do not fully capture the idiosyncrasies of each galaxy. For example, \textbf{m12r} is relatively less massive ($\sim$4$\times$) and has a smaller cold \& dense gas reservoir/lower SFRs compared to the other simulations; \textbf{m12w} (and to a less dramatic extent \textbf{m12c}, though with the same result) exhibits a dramatic lack of gas in its center (within 1 kpc) due to a massive starburst occurring just before the beginning of our analysis, and only near the end of the $\sim 22$ Myr period does the central gas reservoir begin to recover (as a result it does not appear on Fig.~\ref{fig:KS_1deg}, since it does not simultaneously have SFR and gas tracers within 145~pc).  This case is very similar to the gas compaction and inside-out quenching episodes seen in simulations of blue nuggets (that become red nuggets) at higher redshift \citep{Tacchella2016, Tacchella2016a}.  These episodes, however, need not be restricted to high-redshift, as observations with ALMA and in the MaNGA Survey \citep{Lin2020, Brownson2020} have shown similar variations in central star formation efficiency \emph{sans} mergers in green valley galaxies in the local universe.

Several of the galaxies match the CMZ observations simultaneously in $\Sigma_{\rm C\&D}$ and $\Sigma_{\rm SFR}$ \emph{at some point} in this time period, with the galaxies generally exhibiting depletion times on the shorter end of CMZ estimates. Fig.~\ref{fig:KS_1deg} demonstrates this, placing the five galaxies that simultaneously have tracers of star formation and cold \& dense gas in their central $145$~pc on the KS plane (\emph{i.e.}, all but \textbf{m12c} \& \textbf{m12w}), and comparing them with appropriate CMZ observations, and spatially resolved observations of M51 \citep[with 170~pc pixels;][]{Blanc2009} and the Antennae galaxy nuclei \citep[NGC 4038/9, at $\sim$675~pc;][]{Bemis2019}.

Interestingly, both \textbf{m12f}, class (a), and significantly lower-mass \textbf{m12r}, class (b), strongly overlap with the Antennae galaxies KS data, suggesting similarities between merger-induced starbursts and self-driven bursty star formation.  Indeed, simulation work modeling the Antennae interaction by \citet{Renaud2019} and of mergers more generally in the FIRE-2 suite by \citet{Moreno2020} have shown the dramatic effects that these events can have on the gas reservoirs in the central kiloparsec, and subsequent ($\sim$10~Myr later) central SFRs.

Viewed on the KS relation, the simulations exhibit significantly different tracks over 22 Myr, with, \emph{e.g.}, \textbf{m12m} stationary with a nearly constant SFR and cold \& dense gas reservoir, and \textbf{m12b} traveling dramatically across the KS-plane ($\sim$2 dex along either axis). Yet, both of these galaxies at some point overlap with observations of the MW CMZ (which itself has been found to lie on the neutral gas--SFR KS relation, \citealt{Yusef-Zadeh2009}), or parts of the central region of M51.  The SFRs in the class (a) galaxies appear to undergo $\approx$2 starbursts in the 22 Myr window, consistent with feedback/starburst instability timescales discussed in literature \citep{Kruijssen2014a, Benincasa2016, Orr2019}.  None of the galaxies evolve consistently along lines of constant depletion time:  all five shown in Fig.~\ref{fig:KS_1deg} exhibit scatter in their centers about $\tau_{\rm dep} \approx \Sigma_{\rm C\&D}/\Sigma_{\rm SFR} \sim$~100 Myr. Though these central regions can exhibit significant time-variability in their SFRs over few-Myr periods, commensurate with YSO/H$\alpha$ timescales like those of the \citet{Longmore2013} observations, in a time-averaged sense ($\sim$10-100 Myr) they are consistent with the observed KS relation on $\sim$150~pc scales, modulo observational uncertainties \citep{Orr2018}. 

\section{Summary \& Conclusions}

In this letter, we analyzed the central core regions of seven FIRE-2 Milky Way-mass simulated disk galaxies by spatially mapping their SFRs and gas surface densities, and primarily compared them with comparable observations of the Milky Way CMZ.  Our main results are: 

\begin{itemize}
    \item There are two fairly distinct morphological classes of \emph{fiery cores} in this sample, with some galaxies exhibiting very asymmetric/clumpy central gas and star formation distributions (class `a') and others with fairly smooth distributions (class `b').  The intense (concentrated) starbursts in the class (a) cores appear to dramatically alter the gas structures in their centers, whereas the smoother feedback from the star formation cirrus of class (b) cores appear not able to do so.
    
    \item Even in the absence of tuning the initial properties of any of the simulations, we nonetheless find that four of the galaxies analyzed here (\textbf{m12b}, \textbf{m12f}, \textbf{m12m}, \& \textbf{m12r}) are able to match CMZ SFR and gas surface density observations \emph{at some point} in a 22 Myr period.
    
    \item Intriguingly, of the simulated galaxies that simultaneously match MW CMZ gas and SFR observations, half have asymmetric, time-varying gas and SFR distributions (\emph{i.e.}, are in class `a'), whilst the other half are fairly smooth class (b) galactic centers. These results demonstrate that a time-varying model can account for the low star formation efficiency (per mass of dense gas) of the CMZ, and that it is not produced \emph{solely} by some steady state equilibrium. In fact, the presence of bars and the influence of SMBHs may make class (b) galactic centers untenable in reality.
    
\end{itemize}

In all, these simulated galaxies cover a wide range in SFRs and gas surface densities, exhibit marked morphological differences, and some undergo significant changes in the span of only 22 Myr.  The simulations lack SMBHs and (strong) bars, and so we cannot comment on the direct role that either of those would play in shaping and/or regulating the core regions of these galaxies. However, this letter highlights (1) the ability of the FIRE-2 zoom-in simulations to reproduce the ``large scale'' (\emph{i.e.}, 145~pc scale) properties of the CMZ; (2) the marked importance of asymmetric, time-varying (\emph{i.e.}, bursty) star formation and feedback in shaping central galactic regions; (3) that future work with these simulations may help explain how the variety of naturally occurring conditions in central galactic environments arises.

\acknowledgements 
The authors would like to thank Alexander Gurvich, and an anonymous referee, for helpful comments that improved the manuscript.  

CB and HPH gratefully acknowledge support from the National Science Foundation under Award No. 1816715. HPH thanks the LSSTC Data Science Fellowship Program, which is funded by LSSTC, NSF Cybertraining Grant \#1829740, the Brinson Foundation, and the Moore Foundation; his participation in the program has benefited this work. The Flatiron Institute is supported by the Simons Foundation.  AW received support from NASA through ATP grant 80NSSC18K1097 and HST grants GO-14734, AR-15057, AR-15809, and GO-15902 from STScI; the Heising-Simons Foundation; and a Hellman Fellowship. Support for SRL was provided by NASA through Hubble Fellowship grant HST-JF2-51395.001-A awarded by the Space Telescope Science Institute, which is operated by the Association of Universities for Research in Astronomy, Inc., for NASA, under contract NAS5-26555.  RSK acknowledges financial support from the German Research Foundation (DFG) via the Collaborative Research Center (SFB 881, Project-ID 138713538) `The Milky Way System' (subprojects A1, B1, B2, and B8). He also thanks for funding from the Heidelberg Cluster of Excellence STRUCTURES in the framework of Germany's Excellence Strategy (grant EXC-2181/1 - 390900948) and for funding from the European Research Council via the ERC Synergy Grant ECOGAL (grant 855130).




\bibliographystyle{aasjournal}
\bibliography{library,library_hph}



\appendix

\section{(Supplemental) Distribution of In-plane Gas Velocities in Class (a) \& (b) Centers}
\begin{figure*}
	\centering
	\includegraphics[width=0.97\textwidth]{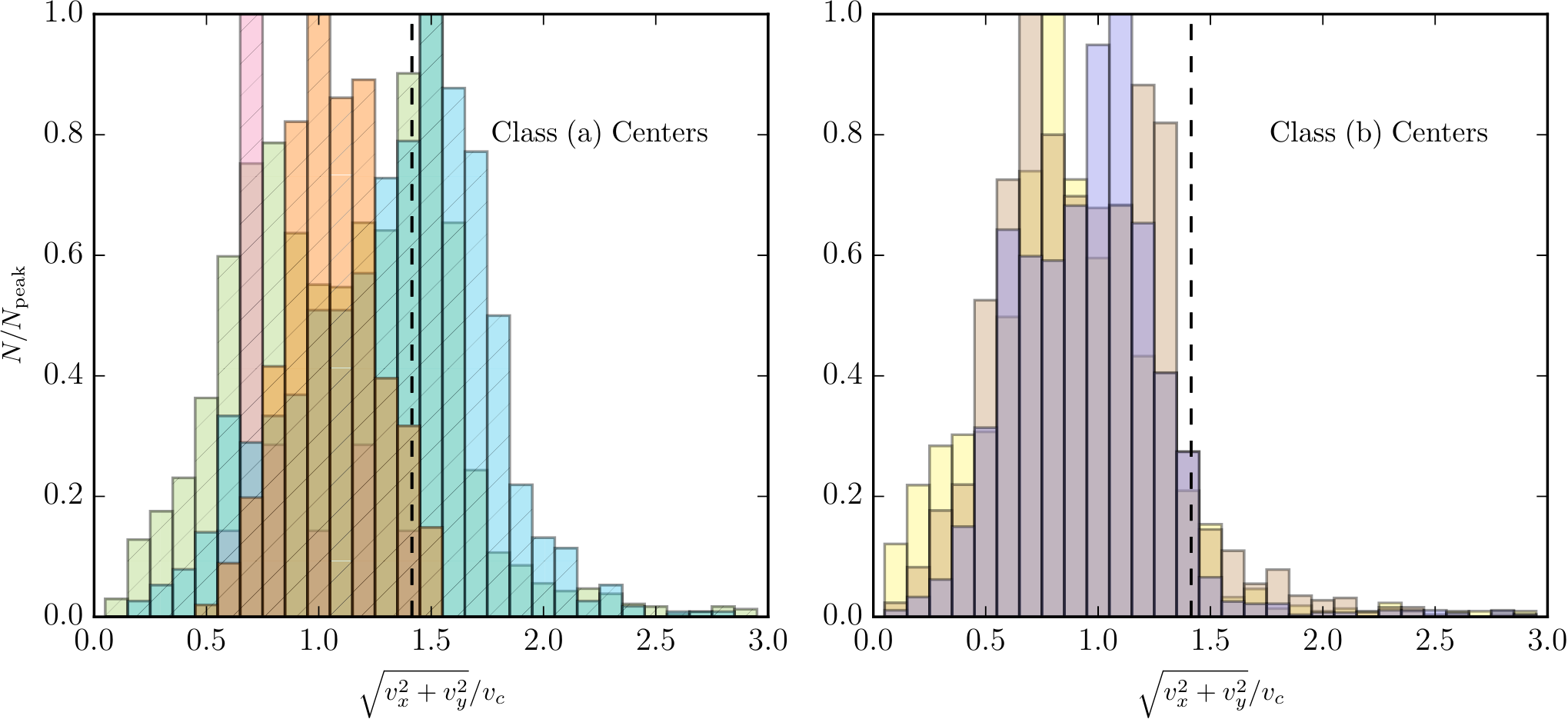}
	\caption{Distribution of in-plane cold \& dense gas velocities (relative to the calculated circular velocity) within the central $R < 500$~pc of the seven FIRE-2 spiral galaxies analyzed in this letter at the final snapshot, split into their respective `classes' (colors and classifications as Fig.~\ref{fig:GASnSFRpanel}). Cold \& dense gas averaged into pixels 10~pc on a side. Dashed vertical line plotted at $\sqrt{2}$ in each panel, noting the classical `escape' velocity, relative to calculated circular velocity (this is not to say that material above $\sqrt{2}v_c$ is unbound here, since the galaxy centers are not point masses, but it does indicate highly non-circular orbits).   \emph{Left panel:} class (a) ``asymmetric'' centers, \textbf{m12b}, \textbf{m12c}, \textbf{m12f}, \& \textbf{m12w}, these galaxy centers have highly variable gas velocity distributions snapshot-to-snapshot (here, \emph{e.g.}, \textbf{m12c} has very little cold \& dense gas, and the distribution from \textbf{m12w} is fairly narrowly centered on $v_c$, but in previous snapshots both more closely resemble the other two broad class (a) distributions shown here), and generally have a significant fraction of material with $\sqrt{v_x^2 + v_y^2} > \sqrt{2} v_c$.  And so, the in-plane gas orbital dynamics are highly variable over short periods ($\Delta t \sim 2.2$ Myr), exhibiting high amounts of non-circularity. \emph{Right panel:} class (b) ``smooth'' centers, \textbf{m12i}, \textbf{m12m}, \& \textbf{m12r}, these galaxies have gas velocity distributions centered close to $v_c$, with very little gas exceeding $\sqrt{2}v_c$.  In short, gas is predominantly moving on circular orbits in these galaxies.  Further, snapshot-to-snapshot, these galaxy centers exhibit much less variation in the shapes of their velocity distributions: they are fairly stable in their orbital dynamics.  \emph{This figure does not appear in the ApJ version of this manuscript, and as such is supplementary material.}}
	\label{fig:orbits}
\end{figure*}

\emph{This appendix does not appear in the ApJ version of this manuscript, and as such is supplementary material.}  In addition to our \emph{by-eye} classification of the seven galaxy centers analyzed here, we briefly investigated the distributions of in-plane gas velocities.  Fig.~\ref{fig:orbits} shows the distributions of gas velocities within $R < 500$~pc for the two classes of galaxy center at the final snapshot of the simulations.  As described in the main text, there is a significant difference between the two classes of galaxy centers, and this extends to this analysis of the velocity distributions: class (a) centers exhibit a large amount of non-circular in-plane gas motion, and are highly variable snapshot-to-snapshot, whereas the class (b) centers have predominantly circular orbital motions, and the distributions are fairly stable snapshot-to-snapshot. Radial motion distributions (not shown) largely tell the same story, class (a) galaxies show (in some cases very large) asymmetry in the distribution of inward and outward flowing gas, and class (b) centers have roughly stable and symmetric distributions (little \emph{net} inflow or outflow).

\label{lastpage}
\end{document}